\documentclass[10pt]{article}
\usepackage[dvips]{graphicx}
\begin{document}

\begin{center}
{\bf Position-dependent stochastic diffusion model of ion channel gating}
 \end{center}

  \begin{center} S. R. Vaccaro  \end{center}
  \begin{center} 
{\em Department of Physics, University of Adelaide, Adelaide, South Australia, 
5005, 
Australia} \\
  \end{center}
{\em E-mail address:}\\
 {\em svaccaro@physics.adelaide.edu.au}\\

{\bf Abstract}
   \begin{quotation}
A position-dependent stochastic diffusion model of gating in ion channels 
is developed by considering the spatial variation of the diffusion coefficient 
between the closed and open states. It is assumed that a sensor which regulates the 
opening of the ion channel experiences Brownian motion in a closed  region  $R_{c}$ 
and a transition region $R_{m}$, where the dynamics  is described by probability 
densities $p_{c}(x,t)$ and $p_{m}(x,t)$ which satisfy interacting Fokker-Planck 
equations with diffusion coefficient $D_{c}(x)=D_{c}\exp(\gamma_{c}x)$ and 
$D_{m}(x)=D_{m} \exp(-\gamma_{m}x)$. The analytical solution of the coupled equations
may be approximated by the lowest frequency relaxation, a short time after the application of a
depolarizing voltage clamp, when $D_{m} \ll D_{c}$ or  the diffusion 
parameter $\gamma_{m}$ is sufficiently large. Thus, an empirical rate
 equation that describes gating transitions may be derived from a 
stochastic diffusion model if there is a large diffusion (or potential) 
barrier between  open and closed states.
\end{quotation}

	

\vspace{0.1in}

{\bf INTRODUCTION}

   Voltage and ligand gated  channels play an important role in initiating 
and modulating the sub-threshold response and the action potential in nerve
 and muscle membranes \cite{hi}. For many years 
the dynamics of the transition between the closed and open states of 
voltage-dependent channels has been described by an empirical rate equation
\begin{equation}
\frac{dP_{o}(t)}{dt} = \alpha - (\alpha + \beta)P_{o}(t),
\label{rate}   \end{equation}
where $\alpha$ and $\beta$ are  opening and closing transition rates 
 and $P_{o}(t)$ is the open state probability \cite{hh}. The dwell-time distribution for the
open state of a nicotinic acetylcholine (nACh) ion channel is also an exponential function
$f_o(t)$ and is associated with the decay of the muscle endplate current \cite{anst}.  If the
ion channel sensor has multiple closed states and an open state, it is assumed that the
 dynamics of the system is described by a master equation. Although the discrete 
state Markov model has been successful in describing ionic and gating currents 
across the membrane, and closed and open dwell-time distributions in ion channels
 \cite{hi,sbkmru,ch}, it does not take account of the Brownian motion of large 
protein molecules in the energy landscape \cite{fsw}.
    
The open or closed state dwell-time distribution $f(t)$ obtained from the patch 
clamp recording of stochastic current pulses in ion channels may be represented 
by a finite sum of exponential functions of time, and for several ion channels, 
$f(t)$ may be approximated by a power law $t^{-p-1}$ for 
intermediate times \cite{mso1,nn}. In order to account for 
multiple relaxation times and the emergence of a power law approximation to the 
dwell-time distribution, both discrete \cite{mso2,la,cj,lieb} and continuous 
\cite{le,ns,gh1,gh2} diffusion models have been proposed, and if it is 
further assumed that there is an increasing barrier height and decreasing energy 
away from the open state, general power laws and a rate-amplitude correlation 
may be derived \cite{va1,va2}. For the Ca-dependent  BK channel, the non-Markovian 
character of the current fluctuations and the dwell-time distribution power law behaviour 
 \cite{fgmsu,mw} may be described by a fractional diffusion model of ion channel
 gating \cite{gh3}.

A numerical solution to a Smoluchowski equation for a voltage-dependent channel 
has shown that a large potential barrier between states ensures that the closed 
state is Markovian with a well-defined escape rate. The gating current has been 
computed for an energy landscape with potential barriers and a spatially 
inhomogeneous diffusion coefficient and is in qualitative agreement with 
experimental data \cite{sqb}.  The objective of the 
paper is to derive an analytical solution of the interacting Fokker-Planck 
equations for a closed region $R_{c}$ and transition region $R_{m}$ in 
response to a depolarizing voltage clamp, and to show that the solution has a single
dominant relaxation time when $D_{m} \ll D_{c}$ or   the 
 diffusion  parameter $\gamma_m$  is sufficiently large.

\vspace{0.1in}

{\bf   STOCHASTIC DIFFUSION MODEL OF ION CHANNEL GATING}

The opening of ligand and voltage activated ion channels  is dependent on the 
conformation of a sensor which is comprised of, in general, several 
macromolecules which may undergo rotation and translation between each surface 
of the membrane \cite{hi,llg}. It is assumed that the sensor experiences Brownian 
motion in a closed state region  $R_{c} (-d_{c} \le x \le 0)$,  and a transition
 region $R_{m} (0 \le x \le d_{m})$  adjacent to the 
open state, with the  dynamics  described by the probability densities 
$p_{c}(x,t)$ and $p_{m}(x,t)$ which satisfy Fokker-Planck (or Smoluchowski) equations 
\cite{kr,ris},
\begin{equation}
\frac{\partial p_c(x,t)}{\partial t} = \frac{\partial}{\partial x} \left[D_c(x) 
\left(\frac{\partial p_c(x,t)}{\partial x} 
+ 
\frac{\partial U_c(x)}{\partial x}  p_c(x,t) \right) \right],
\label{fpc} 
\end{equation}	
\begin{equation}
\frac{\partial p_m(x,t)}{\partial t} = \frac{\partial}{\partial x} \left[D_m(x) 
\left(\frac{\partial p_m(x,t)}{\partial x} 
+ 
\frac{\partial U_m(x)}{\partial x}  p_m(x,t) \right) \right],
\label{fpm}
 \end{equation}
where $U_{c}(x)$ and $U_{m}(x)$ are potential functions.  The diffusion coefficient
 $D_{c}(x)=D_{c}\exp(\gamma_{c}x)$, $D_{m}(x)=D_{m}\exp(-\gamma_m x)$, $\gamma_{c}$
and $\gamma_{m}$ are constants, and either $D_{c}=D_{m}$ or there is a discontinuity
 at the interface between  $R_{c}$ and  $R_{m}$ (see Fig. 1). For Markovian ion channels, 
 the power law approximation to the dwell-time distribution  is dependent 
on the variation in barrier height between closed states \cite{va2}, and therefore we may 
consider the effect of the diffusion  parameters $D_{m}$ and $\gamma_{m}$ in the transition
 region on  the gating dynamics of an ion channel.

\begin{figure*}
\begin{center}
\includegraphics[width=0.7\textwidth]{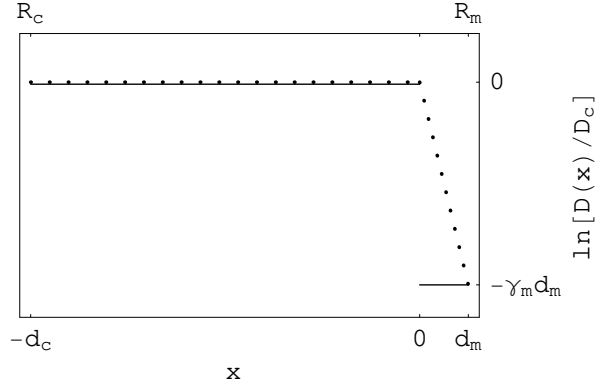}
\caption{
The diffusion coefficient $D(x)$ within  $R_{c}$ and $R_{m}$ may be continuous 
(dotted line) or there may be a discontinuity at  the interface $x=0$ (solid 
line). 
}
\end{center}
\end{figure*}
	The diffusion in the region $R_{c}$ is confined by the inner surface of the 
membrane, and therefore a reflecting boundary is imposed at $x=-d_{c}$
\begin{equation}
\frac{\partial p_c(x,t)}{\partial x} + U_c^{\prime} p_c(x,t) = 0,
\label{bc1} 
\end{equation}
where ${\partial U_c(x)}{\partial x}$ is assumed to be a constant 
$U_c^{\prime}$. Only uni-directional transitions from the closed to the open 
state are considered and therefore $p_{m}(d_{m},t)=0$. It may be assumed that 
the probability current and the probability density are continuous at the interface
 between $R_{m}$ and $R_{c}$
\begin{equation}
 j_c(0,t) = j_m(0,t),
\label{bc2} 
\end{equation}
\begin{equation}
 p_c(0,t) = p_m(0,t).
\label{bc3}
 \end{equation}
The dwell time for each region is $T_{c}= \int_{0}^{\infty} P_c(t) dt$
and $T_{m}= \int_{0}^{\infty} P_m(t) dt$
where the survival probabilities $P_{c}(t) = \int_{-d_c}^0 p_c(x,t)dx$ and 
$P_{m}(t) = \int_{0}^{d_m} p_m (x,t)dx$ \cite{ch}. The ion channel is initially 
in a hyperpolarized state ($P_c(0) = 1$) and hence the initial condition may be specified as
$p_{c}(x,0)= \delta(x)$ and  $p_{m}(x,0)=0$, and we may assume that the ion 
channel is depolarized to a membrane clamp potential of $V=V_{f}$ for which 
$U_{c}(x)$ and $U_{m}(x)$ are independent of $x$.

The relative amplitude of the multiple relaxation times may be determined by 
solving
 Eqs. (\ref{fpc}) and (\ref{fpm}) with the initial and boundary conditions using 
the method of Laplace transforms.  Defining  $z=z_0 \exp(-\gamma_{c}x/2)$,
 $z_0=2/(\gamma_{c} \sqrt{D_{c}})$, $z_{d}=z_0 
\exp(\gamma_{c}d_{c}/2)$, $y=y_0 \exp(\gamma_{m}x/2)$, 
$y_0 =2/(\gamma_{m} \sqrt{D_{m}})$, $y_{d}=y_0 \exp(\gamma_{m}d_{m}/2)$,
 $p_{c}(x,t)=zu_{c}(z,t)$, and $p_{m}(x,t)=yu_{m}(y,t)$, Eqs. (\ref{fpc}) and (\ref{fpm}) may be expressed as 
Bessel differential equations, and it may be shown that the probability that the 
sensor is in the region $R_{c}$ and $R_{m}$ is	
\begin{equation}
P_{cm}(t)= \int_{-d_{c}}^{d_{m}}p(x,t)dx= 
\sum_{i=1}^{\infty}a_{i} \exp(- \omega_{i}t),
 \label{sol1}
\end{equation}  
where $\omega_{i}= \mu_{i}^{2}$ and $\mu_{i}$ is a solution of the characteristic equation
\begin{equation}
\frac{S_{0}(\mu_i,z_0,z_d) S_{1}(\mu_i,y_0,y_d)}
{C_{0}(\mu_i,z_0,z_d) C_{0}(\mu_i,y_d,y_0)} = \sqrt{\frac{D_m}{D_c}},
\label{eigen} \end{equation}
$C_0(\mu_{i},z_0,z_{d})$ and $S_{\nu}(\mu_{i},z_0,z_{d})$ for $\nu = 0$ or $1$  
are defined in terms of Bessel functions of the first and second kinds
\[	
C_0(\mu_{i},z_1,z_2)=J_1(\mu_{i}z_1)Y_0(\mu_{i}z_2)- 
Y_1(\mu_{i}z_1)J_0(\mu_{i}z_2),\]
\[
S_{\nu}(\mu_{i},z_1,z_2) = 
J_{\nu}(\mu_{i}z_1)Y_{\nu}(\mu_{i}z_2)-Y_{\nu}(\mu_{i}z_1)J_{\nu}(\mu_{i}z_2),
\] 
 with similar definitions for the parameters $y_1$ and $y_2$,
\begin{equation}	
a_{i}= \frac{2C_0(\mu_{i},y_{d},y_{d})}
{C_0(\mu_{i},y_{d},y_0)[h_1(\mu_{i})+h_2(\mu_{i})+h_3(\mu_{i})+h_4(\mu_{i})]}, 
\label{amp}
\end{equation}  
\[
h_1(\mu)=\frac{1}{S_0(\mu,z_0,z_{d})} \frac{d[\mu S_0(\mu,z_0,z_{d})]}{d \mu},
\]
\[
h_2(\mu)=\frac{1}{S_1(\mu,y_0,y_{d})}\frac{d[\mu S_1(\mu,y_0,y_{d})]}{d \mu},
\]
\[	
h_3(\mu)=-\frac{1}{C_0(\mu,z_0,z_{d})}\frac{d[\mu C_0(\mu,z_0,z_{d})]}{d \mu},
\] 
\[	
h_4(\mu)=-\frac{1}{C_0(\mu,y_d,y_{0})}\frac{d[\mu C_0(\mu,y_d,y_{0})]}{d \mu}.
\]
 
Adopting a small argument approximation for the Bessel functions \cite{as}, from 
the solution (\ref{sol1})
\begin{equation}	T_{c}= \frac{d_{c}[\exp(\gamma_{m}d_{m})-1]}{D_{m}\gamma_{m} 
}. \label{tc}
\end{equation} 
    From Eqs. (\ref{sol1}) and (\ref{eigen}),  if  $\gamma_{c}$ and $\gamma_{m}$ 
are sufficiently small it may be shown that 
$\omega_1 \approx D_{m}/d_{c}d_{m} \approx 1/T_{c}$ and
\begin{equation}
 \frac{1}{a_1} \approx \frac{1}{2} \frac{\sin \sqrt{\tau_{m}/T_{c}} 
}{\sqrt{\tau_{m}/T_{c}} }[1+\frac{\tau_{m}}{T_{c}} +\frac{\tau_{m}}{T_{c}} 
\frac{\cos^2 \sqrt{\tau_{m}/T_{c}} }{\sin^2 \sqrt{\tau_{m}/T_{c}} } 
(1+\frac{\tau_{c}}{T_{c}})],
\end{equation} 
where $\tau_{m} = d_{m}^2/D_{m}$ and $\tau_{c} = d_{c}^2/D_{c}$.
Therefore $P_{cm}(t)$ may be approximated by the lowest frequency component 
with opening rate $\alpha \approx 1/T_c$  when $\tau_{c} \ll T_{c}$ and $\tau_{m} \ll T_{c}$ or equivalently
\begin{equation}
	\frac{D_{m}}{D_{c}} \ll \frac{d_{m}}{d_{c}} \ll 1, \label{ineq}
\end{equation} 
and is in good agreement with the survival probability of the closed state 
 for a  delayed rectifier K channel, after eliminating the fast closed component with 
a low frequency filter  \cite{clm} (see Fig. 2). The relation $d_{m} \ll d_{c}$ may be
 obtained from the voltage dependence of the mean closed time for an interacting diffusion
regime \cite{gh1} or from the requirement that the probability current in the transition region is 
quasistationary \cite{va2}, and $\tau_{c} \ll T_{c}$ is also satisfied when
there is a large potential barrier in the region $R_m$.
A short time after the application of the voltage clamp, the spatial variation of the 
probability density $p(x,t)$ is approximately linear in the region $R_{m}$ (see Fig. 3) 
and therefore the probability current is constant within the transition region.

\begin{figure*}
\begin{center}
\includegraphics[width=0.7\textwidth]{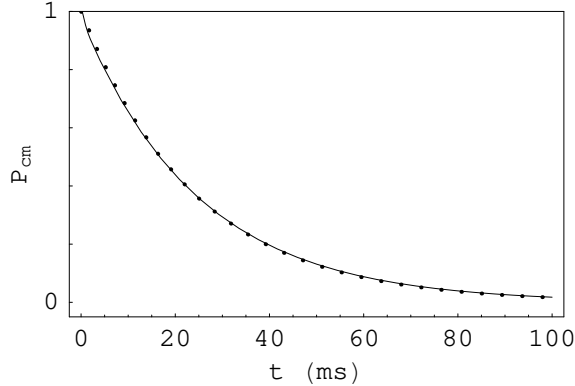}
\caption{
The survival probability of the slow closed state for a delayed rectifier K channel 
\cite{clm} (dotted line)   and the analytical solution 
$P_{cm}(t)$ (solid line) where $d_{m}/d_{c}=0.15$, $D_{m}/D_{c}=0.0225$, 
$\tau_c = 3.4$ ms, $\gamma_{c},\gamma_{m} \rightarrow 0$,
$a_{i}=(0.977,0.130,-0.103...)$ and $\omega_{i}=(0.040,2.26,3.62...)$
}
\end{center}
\end{figure*}

By assuming that $D_{m}(x)=D_{c}(x)=D$ and $d_{m} \ll d_{c}$, it follows that 
$T_{c}=\tau_{c}d_{m}/d_{c} \ll \tau_{c}$ and the dwell-time distribution for 
intermediate times may be described by a power law (see Fig. 4)  \cite{mso2,cj,gh1}, as 
observed in several types of ion channels for the closed states accessible from the open state 
during a depolarizing patch clamp. However, it should be noted that the closed states associated
with a power-law approximation to the dwell-time distribution are, generally, not the same as 
 those in the activation sequence \cite{zha}.
\begin{figure*}
\begin{center}
\includegraphics[width=0.7\textwidth]{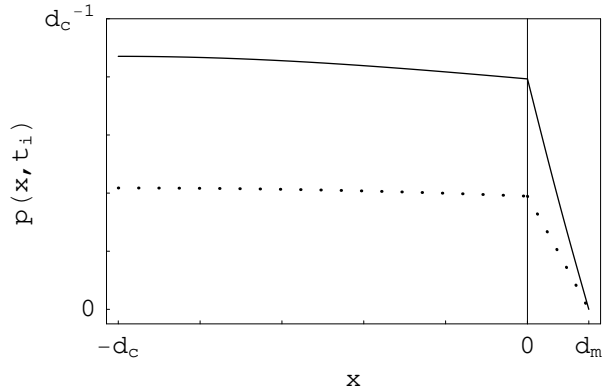}
\caption{
The probablility density $p(x,t)$ in the region $R_{c}(-d_{c} \le x \le 0)$ and 
$R_{m}(0 \le x \le d_{m})$ for $t_1 = 2$ ms (solid line) and $t_2 = 20$ ms (dotted line) 
where $d_{m}/d_{c}=0.15$, $D_{m}/D_{c}=0.0225$, $\tau_c = 3.4$ ms, 
$\gamma_{c}$, $\gamma_{m} \rightarrow 0$.
}
\end{center}
\end{figure*}

If $D_{m}=D_{c}$, $\gamma_{c}d_{c} \ll 1$ and $\gamma_{m}d_{m} \gg 1$, adopting 
a small argument approximation in $R_{m}$ and large argument approximation in 
$R_{c}$  \cite{as}, it may be shown from Eqs. (\ref{sol1})  and (\ref{eigen})  that 
\begin{equation}
\frac{1}{\omega_1} \approx
\frac{d_{c}[\exp(\gamma_{m}d_{m})-1]}{D_{m}\gamma_{m}} \approx T_{c},
\label{cond2}
\end{equation} 
 \begin{equation}
\frac{1}{a_1} \approx 1 + \frac{\gamma_{m}d_{c}}{2[\exp(\gamma_{m}d_{m})-1]},
\end{equation} 
 and thus $a_1 \approx 1$ and $a_{i} \approx 0$ for $i>1$ when 
\begin{equation}
\frac{\gamma_{m}d_{m}}{\exp(\gamma_{m}d_{m})-1} \ll \frac{d_{m}}{d_{c}},
\end{equation} 
or from Eq. (\ref{tc}), $\tau_{c} \ll T_{c}$. 
 Therefore, when $\gamma_{m}$ is sufficiently large,   
$P_{cm}(t) \approx \exp(- \omega_{1}t)$  and in agreement with the data from  a 
 delayed rectifier K channel \cite{clm} (see Fig. 5)

\begin{figure*}
\begin{center}
\includegraphics[width=0.7\textwidth]{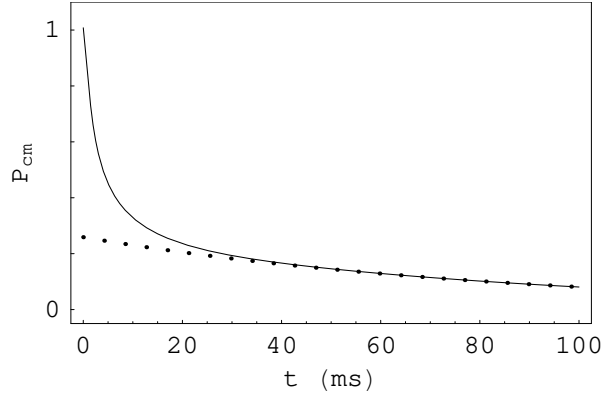}
\caption{
The survival probability for the closed state $P_{cm}(t)$ (solid line) and the 
lowest frequency component $a_1 \exp(-\omega_1 t)$ (dotted line) where 
$d_{m}/d_{c}=0.15$, $D_{c}=D_{m}$, $T_c = 24$ ms, $\gamma_{c}$, $\gamma_{m} \rightarrow 0$, $a_i =(0.259,0.245,0.218...)$ and $\omega_{i}=(0.012,0.105,0.292...)$.
}
\end{center}
\end{figure*}

{\bf DISCUSSION}

Gating in voltage or ligand activated ion channels is regulated, in general, by several
 macromolecules which experience Brownian motion in the closed and open regions, where  
the dynamics may be described by probability densities which satisfy 
interacting Fokker-Planck equations. We have shown that 
 a single dominant relaxation time may be derived from a position-dependent 
stochastic diffusion model when there is a discontinuity in the diffusion coefficient at 
the interface between the regions $R_{m}$ and $R_{c}$ with $D_{m} \ll D_{c}$, 
and the width of the transition region ($d_{m}$) is much less than  the width of the closed 
region ($d_{c}$). These conditions ensure that $\tau_{c} \ll T_{c}$ and 
$\tau_{m} \ll T_{c}$, and therefore the Brownian motion in the 
closed and transition regions may be described as quasi-stationary.
The  small value of $d_m$ is consistent with recent experimental data that 
indicates that each S4 sensor has a translation of the order of
 6 $\AA$ across a focussed electric field \cite{ah}.  If 
$D_m(x) = D_{m} \exp(-\gamma_{m}x)$, the response of the system to a 
depolarizing voltage clamp may also be approximated by the lowest frequency 
relaxation when the diffusion parameter $\gamma_{m}$ is sufficiently large.

 If the opening of the ion channel is determined by $m$ identical and 
independent subunits and the conductance of the channel is expressed as 
$g \propto P_{o}(t)^{m}$ \cite{hh}, a rate equation may be derived for each 
subunit when there is a large diffusion or potential barrier between the 
closed and open configurations of each sensor molecule. When the closed state
 dwell-time distribution obtained from a patch clamp recording has a finite
 number of relaxation times, the  closed states may be represented as energy wells
 between potential or diffusion barriers within  an energy landscape,
and the resulting  system of interacting Fokker-Planck equations may be 
approximated by a Markovian master equation.  

\begin{figure*}
\begin{center}
\includegraphics[width=0.7\textwidth]{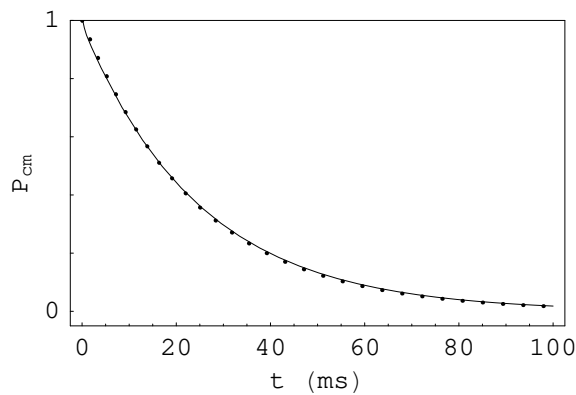}
\caption{
The survival probability of the slow closed state for a delayed rectifier K channel
 \cite{clm} (dotted line) and the analytical solution 
 $P_{cm}(t)$ (solid line) where $d_{m}/d_{c}=0.15$, $\gamma_{c} \rightarrow 0$,
 $\gamma_{m} d_m = 6$,$\tau_c = 2.2$ ms, $a_{i}=(0.984,0.055,-0.030...)$ 
and $\omega_{i}=(0.04,3.5,7.48...)$.
}
\end{center}
\end{figure*}

 \end{document}